\documentstyle[amssymb,preprint,prl,aps]{revtex}

\begin{document}
\title{Insights into the superdiffusive dynamics through collision statistics in
periodic Lorentz gas and Sinai billiard }
\author{$^{1}$Valery B. Kokshenev and $^{2}$Eduardo Vicentini}
\address{$^{1}${\it Departamento de Fisica, Universidade Federal de Minas Gerais,
Caixa Postal 702, 30123-970, Belo Horizonte, Minas Gerais, Brazil}\\
$^{2}${\it Departamento de F\'{i}sica, Universidade Estadual do Centro
Oeste, Caixa Postal 730, CEP 85010-990,} {\it Guarapuara, PR, Brazil\bigskip 
}\\
\today }
\maketitle

\begin{abstract}
\leftskip 54.8pt \rightskip 54.8pt

We report on the stationary dynamics in classical Sinai billiard (SB)
corresponding to the unit cell of the periodic Lorentz gas (LG) formed by
square lattice of length $L$ and dispersing circles of radius $R$ placed in
the center of unit cell. Dynamic correlation effects for classical
particles, initially distributed by random way, are considered within the
scope of deterministic and stochastic descriptions. A temporal analysis of
elastic reflections from the SB square walls and circle obstacles is given
for distinct geometries in terms of the wall-collision and the
circle-collision distributions. Late-time steady dynamic regimes are
explicit in the diffusion exponent $z(R)$, which plays a role of the
order-disorder crossover dynamical parameter. The ballistic ($z_{0}=1$)
ordered motion in the square lattice ($R=0$) switches to the superdiffusion
regime with $z_{1}=1.5$, which is geometry-independent when $R<L\sqrt{2}/4$.
This observed universal dynamics is shown to arise from long-distance
particle jumps along the diagonal and nondiagonal Bleher corridors in the LG
with the infinite horizon geometry. In the corresponding SB, this universal
regime is caused by the long-time wall-collision memory effects attributed
to the bouncing-ball orbits. The crossover nonuniversal behavior with $%
1.5<z<2$ is due to geometry with $L\sqrt{2}/4\leq R<L/2$, when only the
nondiagonal corridors remain open. All the free-motion corridors are closed
in LG with finite horizon ($R\geq L/2$) and the interplay between square and
circle geometries results in the chaotic dynamics ensured by the normal
Brownian diffusion ($z_{2}=2$) and by the normal Gaussian distribution of
collisions. PACs: 05.45.Gg , 05.40.Fb, 45.50.Tn. Key words: Sinai billiard,
Lorentz gas, collision statistics, anomalous diffusion, order-disorder
crossover, chaos. Corresponding author Valery B. Kokshenev,
valery@fisica.ufmg.br
\end{abstract}

\newpage

\section{INTRODUCTION}

A mathematical exploration of two-dimensional Lorentz gas (LG) with periodic
configuration of scatterers, related to dispersed billiards, was initiated%
\cite{BS81} by Bunimovich and Sinai. The process of establishing equilibrium
dynamics for randomly distributed, noninteracting classical particles driven
by elastic collisions with the billiard walls and the scatterer of fixed
geometry are commonly discussed in terms of the particle velocity-velocity
and/or displacement-displacement correlation functions, which are related to
a {\em diffusion coefficient} through the Einstein-Green-Kubo formula (see
e.g. Ref.\cite{Ble92}). Besides the ergodicity, the entropy, the Lyapunov
exponent, the mixing property, and among other interesting physical
observables in LG with finite and infinite horizons, the enhanced diffusion
studied numerically through the collision distribution function, has been
expressed as of great importance in Ref.\cite{GG94}. This stimulated our
subsequent investigations of the collision statistics for nonescape
particles observed through their survival probability in {\em weakly open}
chaotic (Sinai billiard\cite{Sin70}, (SB))\cite{KN00} and non-chaotic
(circle and square billiards)\cite{VK01} classical systems. Remarkably,
collision statistics provided new insight into the delicate mathematical
problem of the interplay between regular and irregular segments of the
billiard boundary, as has been demonstrated for the case of the so-called
almost-integrable systems, presented by open\cite{KV03} and closed\cite{RC02}
rational polygons. In the current study we focus on the stationary
superdiffusion dynamics in the closed SB. Analysis is given through the
billiard-wall and the scatterer-disk random collision statistics and is
based on the dynamical correspondence between SB and LG. Some of the
findings of this study were preliminary communicated in Ref.\cite{RC02}.

The dynamics of particles (of unit mass and unit velocity) moving in
two-dimensional closed region (billiard table) and dispersed by obstacles is
governed by billiard-boundary geometry. The SB\ with the square-wall table
of length $L$ and the disk with radius $R$ can be formally treated as the
unit cell of a periodic LG, which is the two-dimensional periodic crystal
formed by a regular set of circular scatterers (of radius $R$) centered at
distances $L$. This implies that SB and LG are dynamically equivalent
classical systems. This statement is justified by a step-to-step
correspondence that can be established between a certain orbit in the closed
SB and the corresponding trajectory in LG. For the case of the low-density
LG, given by geometry with $R<L/2$, this correspondence is exemplified in
inset A in {\bf Fig. 1}. As seen for this LG configuration, there exist
trajectories in which particles never collide with scatterers. Those
particles, which move through such unbounded trajectories, have therefore an 
{\em infinite horizon}\cite{Ble92,GG94}. In turn, the free-motion
trajectories belong to infinite corridors\cite{Ble92}. As shown by Bleher%
\cite{Ble92}, the principal corridors are open for small disk radii limited
by $0<R<L\sqrt{2}/4$. This is illustrated by the diagonal and nondiagonal
corridors in inset B in Fig.1. When the scatterer radius achieves the
magnitude of $L\sqrt{2}/4$, the last nondiagonal corridor closes. With
further increasing the radius, the principal corridors disappear at $L/2$
and scatterers start to overlap lattice cells. The particle trajectories
become bounded thus having a {\em finite horizon} (see inset C in Fig.1).

In the infinite-horizon LG, the enhanced-diffusion motion regime was
theoretically predicted in Ref.\cite{Ble92}, through the asymptotic
statistical behavior of the late-time particle displacements. But no
description for the evolution of motion regimes with geometry was given (see
also Ref.\cite{Dah96}). We therefore reformulate the problem of random
particle displacements into that of random collisions. This yields a
description of the late-time stationary dynamics in LG, as well as in the
corresponding SB, through the {\em diffusion exponent} $z(R)$, which is
continuous with $R$. The paper is organized as follows. In Sec. II we
develop the billiard collision statistics based on the alternative
deterministic and stochastic approaches. Also, we introduce the collision
distribution function given in terms of the dynamic observables, which are
available in simulation of SB\ with finite and infinite horizon geometries
in the corresponding LG. Conclusion is drawn in Sec. III.

\section{COLLISION STATISTICS}

\subsection{Distribution Function and Dynamic Characteristics}

The trajectories of classical particles of unit mass moving with unit
velocities in SB (or in LG) are preserved by the Liouville measure\cite
{Cher88}, namely 
\begin{equation}
d\mu ({\bf x})=\frac{1}{2\pi A}dxdy\,d\theta \text{.}  \label{Lm}
\end{equation}
This is introduced in the phase space through the billiard table area $A$,
given by the variable scatterer radius $R$ and the fixed boundary side $L$.
The coordinate set ${\bf x}=(x,y,\theta )$ includes the particle position
and the velocity launching angle $\theta =[0,2\pi ]$, which is counted of
the $x$-axis of the square billiard table. The {\em mean collision time} $%
\tau _{c}(R)$, which is due to the two consequent elastic random collisions
with the boundary wall or with the scatterer, are respectively defined\cite
{Cher88} by 
\begin{equation}
\tau _{c}^{(w)}(R)=\frac{\pi A}{P_{w}}\text{ or }\tau _{c}^{(s)}(R)=\frac{%
\pi A}{P_{s}}\text{.}  \label{tau-c}
\end{equation}
Here $P_{w}$ $=4L$ and $P_{s}=2\pi R$ are corresponding collision perimeters
of the table with accessible area $A$ $=L^{2}-\pi R^{2}$ ( for $R<L/2$).

The {\em collision} {\em distribution function} $D(n,t)$ is a probability of
a particle to collide $n$ times with the fixed billiard boundary within a
time $t$ (for rigorous definition see {\it e.g.} Ref.\cite{GG94}). Also, it
can be introduced through the billiard mean collision-number equations,
namely

\begin{equation}
n_{c}(t)=<n({\bf x,}t)>_{c}=\int n({\bf x,}t)\,d\mu ({\bf x}%
)=\int_{0}^{\infty }n\,D(n,t)dn=\frac{t}{\tau _{c}}\text{,}  \label{n1}
\end{equation}
where ${\bf x}$ stands for the boundary (wall and/or scatterer) position set
and $\tau _{c}(R)$ is given in Eq.(\ref{tau-c}). This results in

\begin{equation}
D(n,t)=\left| \frac{d\mu [{\bf x}(n,t)]}{dn}\right| \text{,}  \label{D-R}
\end{equation}
where the Liouville measure, defined in Eq. (\ref{Lm}), is given by the
inverse function to $n({\bf x,}t)$.

The distribution (\ref{D-R}) provides a rich information on the
boundary-memory effects in chaotic SB with a fixed geometry. In other words,
dynamic correlation effects can be characterized by the nonzero central
moments of order $m=2,4...$ defined for the random numbers $n$ as 
\begin{equation}
\Delta ^{m}n_{c}(t)=\int_{0}^{\infty }[n-n_{c}(t)]^{m}\,D(n,t)dn\text{. }
\label{delta-n-k}
\end{equation}
$\Delta ^{m}n_{c}(t)$ describes $m$-order deviation from the mean collision
number $n_{c}(t)$ given in Eq.(\ref{n1}). Basically, we focus on low-order
dynamic correlation effects presented by the {\em variance} of the random
collision numbers $n$%
\begin{equation}
\Delta ^{2}n_{c}(t)=\frac{2}{\tau _{c}^{2}}\langle \Delta ^{2}r\rangle _{c}%
\text{,}  \label{delta-n}
\end{equation}
expressed through the variance for particle displacements $\langle \Delta
^{2}r\rangle _{c}$. This fundamental equation will be deduced below using
the orbit-trajectory correspondence visualized in Fig. 1. Also, Eq.(\ref
{delta-n}) permits one to employ the well known temporal
variance-displacement equation (see {\it e.g.} Ref.\cite{MK00})

\begin{equation}
\langle \Delta ^{2}r\rangle _{c}\backsim \ell _{c}^{2}\left( \frac{t}{\tau
_{c}}\right) ^{2/z}\text{, with }t\gg \tau _{c}\text{.}  \label{delta-r}
\end{equation}
In this way, the {\em diffusion exponent} $z(R)$ introduces a description
for distinct stationary regimes in billiards with different $R$. For
particles of unit velocity, the mean free path $\ell _{c}=\tau _{c}$ can be
specified for the wall and the scatterer collisions with the help of Eq.(\ref
{tau-c}).

The {\em wall-collision} and the {\em scatterer-collision} statistics is
given by the corresponding distribution functions $D^{(w)}(n,t)$ and $%
D^{(s)}(n,t)$. These are defined by the equations $n_{c}^{(w)}(t)=t/\tau
_{c}^{(w)}$ and $n_{c}^{(s)}(t)=t/\tau _{c}^{(s)}$, which extend Eq.(\ref{n1}%
), as well as Eqs. (\ref{D-R}) and (\ref{delta-n-k}). Furthermore, from the
same equations one has $\tau _{c}^{(w)}n_{c}^{(w)}=\tau
_{c}^{(s)}n_{c}^{(s)}=t$ that, with accounting for the total number of
collisions $n_{c}(t)=n_{c}^{(w)}+n_{c}^{(s)}$ with $\tau _{c}n_{c}=t$ ,
results in the overall mean collision frequency, namely

\begin{equation}
\frac{1}{\tau _{c}}=\frac{1}{\tau _{c}^{(w)}}+\frac{1}{\tau _{c}^{(s)}}=%
\frac{P_{w}+P_{s}}{\pi A}\text{ .}  \label{1/tau-c}
\end{equation}
Here the mean wall ($\tau _{c}^{(w)}$) and scatterer ($\tau _{c}^{(s)}$)
collision times are defined in Eq.(\ref{tau-c}).

\subsection{Wall Collisions in Square Billiard}

If one ignores spliting effects caused by $\pi /2$-angle vertices\cite{KV03}%
, particle in the closed square billiard are subjected to an ordered orbit
motion driven by billiard walls. This implies that the velocity launching
angle $\theta $ is the integral of motion for a given orbit, which can be
therefore presented by a straight-line trajectory in the corresponding LG
(see inset D in Fig. 1). A whole number of intersections of this line with
the unit-cell boundaries, encounted in $x$ and $y$ directions in a time $t$,
corresponds to the following square-billiard, {\em wall-collision }number,
namely\cite{MA} 
\begin{equation}
n_{0}^{(w)}(\theta ,t)=\frac{t}{L}\left( \cos \theta +\sin \theta \right) 
\text{, }  \label{n-teta}
\end{equation}
established for a given orbit by the velocity angle $\theta $. In view of
the point symmetry of square lattice, the angle domain is reduced to $0\leq
\theta \leq \pi /4$. Extending Eq.(\ref{n1}) for the wall-ordered motion,
the corresponding characteristic time $t_{c}^{(w)}(\theta )$ follows from
equation $n_{0}^{(w)}(\theta ,t)=t/t_{c}^{(w)}(\theta )$. This gives 
\begin{equation}
\frac{1}{t_{c}^{(w)}(\theta )}=\frac{\cos \theta +\sin \theta }{L}\text{,}
\label{f-teta}
\end{equation}
obtained with the help of Eq.(\ref{n-teta}). Consequently, the mean
collision number is 
\begin{equation}
n_{c0}(t)\equiv \left\langle n_{0}^{(w)}(\theta ,t)\right\rangle _{c}=\frac{%
4t}{\pi A}\int\limits_{0}^{\pi /4}\frac{\cos \theta +\sin \theta }{L}d\theta
\int\limits_{0}^{A}dxdy=\frac{4t}{\pi L}=\frac{t}{\tau _{c0}^{(w)}}\text{.}
\label{n-c-0}
\end{equation}
Here the averaging procedure is elaborated over all equivalent angles $%
\theta $. In turn, Eq.(\ref{n-c-0}) defines the mean collision time $\tau
_{c0}$ that agrees with Eq.(\ref{tau-c}), where $A=L^{2}$ and $P_{w}=4L$.
The variance for the random wall-collision number is 
\begin{equation}
\Delta ^{2}n_{c0}(t)=\left\langle [n_{0}^{(w)}(\theta
,t)-n_{c0}(t)]^{2}\right\rangle _{c0}=\left( \frac{\pi ^{2}}{16}+\frac{\pi }{%
8}-1\right) n_{c0}^{2}\text{,}  \label{var}
\end{equation}
with $n_{c0}$ is obtained in Eq.(\ref{n-c-0}). The function $\theta
(n,t)=0.5\arcsin \left[ \left( 4\tau _{c0}n/\pi t\right) ^{2}-1\right] $,
inverse to the function $n(\theta ,t)$ given in Eq.(\ref{n-teta}), results
in the collision distribution function $D_{0}(n,t)=4\pi ^{-1}\left| \partial
\theta (n,t)/\partial n\right| $ defined in Eq. (\ref{D-R}). A
straightforward estimation for the wall-collision distribution yields 
\begin{eqnarray}
D_{0}^{(w)}(n,t) &=&\frac{16}{\pi ^{2}n_{c0}\sqrt{2}}\sin ^{-1}\left( \frac{%
\pi }{4}-\frac{1}{2}acr\sin \left[ (\frac{4}{\pi }\frac{n}{n_{c0}}%
)^{2}-1\right] \right) \text{,}  \nonumber \\
\text{ for }\pi /4 &<&n/n_{c0}<\pi \sqrt{2}/4\text{, otherwise }%
D_{0}^{(w)}(n,t)=0\text{.}  \label{D-0}
\end{eqnarray}
One can verify that Eqs. (\ref{n-c-0}), (\ref{var}) and (\ref{D-0}) are
selfconsistent hence obey Eqs.(\ref{n1}) and (\ref{delta-n-k}). As seen from
Eqs. (\ref{delta-r}) and (\ref{var}), the dynamical diffusion exponent $%
z_{0}=1$, {\it i.e.}, the dynamic regime in the square billiard is
ballistic. Meantime, ballistic trajectories were employed to deduce the
square-billiard distribution (\ref{D-0}).

In {\bf Fig. 2} the wall-collision distribution, which is predicted by Eq.(%
\ref{D-0}), is compared with that simulated in square billiard at
observation time $t_{obs}=100\tau _{c0}$ (for experimental details, see Ref.%
\cite{VK01}). Analysis for whole-scale temporal evolution was also performed%
\cite{Tese}. We therefore infer that the simulated root-mean-square
deviation for collision numbers $\sqrt{\Delta ^{2}n_{c0}(t)}$ agrees with
the stationary prediction given in Eq.(\ref{var}), starting with times $%
t_{obs}\gtrsim 30\tau _{c0}$ (see the inset in Fig.2).

\subsection{Random Walks in Lorentz Gas Lattice}

\subsubsection{Infinite Horizon Geometry}

In the case of the scatterer radii $0<R<L/2$, besides a free motion along
the open corridors in the phase space of the chaotic SB, particles are
dispersed by disks and in this way are involved in diffusive motion. Let us
describe an evolution of a given trajectory, moving in the equivalent
square-lattice LG, by random walking (see inset A in Fig. 1).

In a time $t$, a walker scattered by disks indicates random steps: $%
n(t)=s_{x}^{+}+s_{x}^{-}+s_{y}^{+}+s_{y}^{-}$. These steps are given by
interceptions of a trajectory with the LG unit cells, that corresponds to
reflections from the SB walls. More precisely, the walker under
consideration does $s_{x}^{+}$ steps to the right, $s_{x}^{-}$ steps to the
left in the $x$-direction as well as $s_{y}^{+}$ steps to the down, and $%
s_{y}^{+}$ steps to the up in the $y$-direction. The resultant particle
displacement is therefore given by 
\begin{equation}
\Delta {\bf r(}t{\bf )}=(s_{x}^{+}-s_{x}^{-})\ell _{x}{\bf e}%
_{x}+(s_{y}^{+}-s_{y}^{-})\ell _{y}{\bf e}_{y}\text{.}  \label{delta-r-n}
\end{equation}
Here $\ell _{x}$ and $\ell _{y}$ are projections of the one-step,
free-motion displacement that occurs between the two consequent
intersections. For random walking in the $x$-direction, $%
<s_{x}^{+}-s_{x}^{-}>_{c}=0$ within the same approximation, and thus $%
<\Delta {\bf r}>_{c}=0$. For the mean squared displacement $<\Delta {\bf r}%
\Delta {\bf r}>_{c}$, one has

\begin{equation}
<\Delta ^{2}r>_{c}=2\ell _{c}^{2}\left[
<s_{i}^{2}>_{c}-<s_{i}s_{k}>_{c}(1-\delta _{ik})\right] ,\text{ }
\label{delta-r2}
\end{equation}
with the help of Eq.(\ref{delta-r-n}), where $\delta _{ik}$ is the Kronecker
symbol. Here the mean $<\ell _{x}^{2}$+$\ell _{y}^{2}>_{c}=2<\ell
_{x}^{2}>_{c}=\ell _{c}^{2}$ is estimated in the isotropic approximation.
The introduced indices $i,k=1,2,3,4$ count distinct random steps: $%
s_{1}=s_{x}^{+}$, $s_{2}=s_{x}^{-},$ $s_{3}=s_{y}^{+}$, and $s_{4}=s_{y}^{-}$%
, which are dynamically equivalent: $<s_{i}>_{c}=<s_{k}>_{c}$, for any $%
i\neq k$. This observation permits one to describe the random-wall
collisions through the random-walk steps, on the basis of the relation $%
n=\sum\limits_{i=1}^{4}s_{i}$. This results in the mean $n_{c}$ and the
variance $\Delta ^{2}n_{c}$, namely

\begin{equation}
n_{c}=\sum\limits_{i=1}^{4}\langle s_{i}\rangle _{c}=4\langle s_{i}\rangle
_{c}\text{, }\Delta ^{2}n_{c}=\sum\limits_{i=1}^{4}\langle \Delta
^{2}s_{i}\rangle _{c}+2\sum\limits_{i>k}\langle \Delta s_{i}\Delta
s_{k}\rangle _{c}\text{, }  \label{n-cR-s}
\end{equation}
introduced by the random step fluctuations $\Delta s_{i}=s_{i}-\langle
s_{i}\rangle $, discussed above in Eqs. (\ref{n1}) and (\ref{delta-n}).

In the stochastic approximation, the step fluctuations are independent and
the last term in Eq.(\ref{n-cR-s}) is null. Moreover, one can see that the
term $\langle \Delta ^{2}s_{i}\rangle _{c}=(<s_{i}^{2}>_{c}-<s_{i}>_{c}^{2})$
is the same as in the square brackets in Eq.(\ref{delta-r2}). This leads to
Eq.(\ref{delta-n}), that in combination with Eq.(\ref{delta-r}) provides the
desired relation for the variance for collision deviations, namely

\begin{equation}
\Delta ^{2}n_{c}(t)\backsim \left( \frac{t}{\tau _{c}(R)}\right) ^{2/z},
\label{delta-n-z}
\end{equation}
where the characteristic billiard collision time $\tau _{c}(R)$ is given in
Eq.(\ref{tau-c}).

On the basis of Eq.(\ref{delta-n-z}), we have elaborated a numerical
statistical analysis for the wall collisions in SB with $0<R<L/2$. As seen
from the upper inset in {\bf Fig. 3}, the standard deviation indicates two
superdiffusion motion regimes, which are well distinguished through the
dynamic exponent $z(R)<2$. For small scatterers, the $R$-independent and,
therefore, universal regime is manifested by $z_{1}=1.50\pm 0.05$ (shown by
closed squares in Fig.3). Above a crossover radius $\thickapprox 0.35L$, the
diffusion exponent continuously increase with $R$ from $1.5$ to $2$ (shown
by open squares in Fig.3). A crossover from the universal to a transient
dynamics starts at $R_{1}=\sqrt{2}L/4$ $\thickapprox 0.35L$, the point where
a rearrangement of the LG lattice occurs with closing of the diagonal Bleher
corridors (shown in inset B in Fig. 1).

We focus on the late-time asymptotic collision behavior. The observed
distributions have been therefore reestimated in the reduced coordinates
proposed in Ref.\cite{GG94}. These coordinates can be formally introduced
here by the relations

\begin{equation}
\widetilde{D}(\widetilde{n})=\sqrt{\Delta ^{2}n_{c}(t)}D[n(\widetilde{n}%
,t),t]\text{ and }\widetilde{n}(n,t)=\frac{n-n_{c}(t)}{\sqrt{\Delta
^{2}n_{c}(t)}}\text{,}  \label{D-red}
\end{equation}
where $n(\widetilde{n},t)$ stands for the inverse function to $\widetilde{n}%
(n,t)$. The distribution $D(n,t)$, the mean $n_{c}(t)$, and the variance $%
\Delta ^{2}n_{c}(t)$ are given in Eqs.(\ref{D-R}), (\ref{n1}) and (\ref
{delta-n-z}), respectively. Physically, the reduced collision number $%
\widetilde{n}(n,t)$, as well as its distribution $\widetilde{D}(\widetilde{n}%
)$, is expected to expose a stationary behavior when the late-time
observation condition $t_{obs}\gg \tau _{c}(R)$ is satisfied (see also Eq.(%
\ref{delta-r})). As follows from our temporal analysis shown in the insets
in Fig. 3, the superdiffusion dynamic regime becomes steady starting from
the observation times $t_{obs}^{(\exp )}\gtrsim 50\tau _{c}$. The same can
be referred to the steady collision distributions exemplified in {\bf Fig. 4}%
.

As seen from Fig. 4, both the kinds of collision distributions are similar.
With growth of the scatterer radius, they change their form from
characteristic for the ordered motion (illustrated in Fig.2 ) to that which
tends to the disordered Gaussian motion, shown by solid lines in Fig. 4.
Furthermore, according to simulation studies carried out within the domain $%
0.05<R/L<0.35$, the late-time distribution functions fall down into the
coinciding curves $\widetilde{D}_{1}(\widetilde{n})$, attributed to the
universal superdiffusion regime introduced above by the exponent $%
z_{1}=1.50\pm 0.05$. This regime is ensured by the open diagonal and
nondiagonal corridors, that makes plausible to adopt that the mean collision
time is caused mostly by the wall reflections. This implies that $\tau
_{c1}\thickapprox \tau _{c1}^{(w)}<\tau _{c1}^{(s)}$ and thus $%
n_{c1}^{(w)}>n_{c1}^{(s)}$ estimated for $R<R_{1}$ (see Eq.(\ref{1/tau-c})).
Consequently, the mean collision number $n_{c1}=<n_{1}({\bf x}%
,t)>_{c}\thickapprox t/\tau _{c1}^{(w)}$ , defined in Eq.(\ref{n1}), can be
specified through random walks, namely

\begin{equation}
n_{1}({\bf x,}t)=\sum_{i}^{n_{c1}^{(s)}}\frac{r_{i}}{t_{c1}^{(w)}(\theta
_{i})}\thickapprox \frac{r}{\tau _{c1}^{(w)}}\text{ , }  \label{n1-ran}
\end{equation}
with $r=t=\sqrt{x^{2}+y^{2}}$ defined by the position set ${\bf x}%
=(x,y,\theta )$. Eq.(\ref{n1-ran}) describes the {\em scatterer-scatterer}
collision, which occurs in the periodic LG during time $t$ between two
scatterers connected by the vector ${\bf x}$, with $\sum_{i}{\bf x}_{i}={\bf %
x}$; $t_{c}^{(w)}(\theta _{i})$ is defined by Eq.(\ref{f-teta}) for a given
linear trajectory $i$, which corresponds to the two consequent wall-to-wall
collisions.

On the other hand, the superdiffusion regime under discussion can be treated
through the two-dimensional L\'{e}vy jumps, which occur between two
scatterers of distance $r$. Therefore, the universal regime can be
additionally characterized by the waiting-time probability distribution
function

\begin{equation}
\Psi (r,t)=\Lambda _{s}(r)\delta (r-t)\text{ with }\Lambda _{s}(r)\varpropto
r^{\frac{2}{z}-5}\text{,}  \label{psi}
\end{equation}
with $1<z<2$ and the jump-length distribution $\Lambda _{s}(r)$ function
presented in the long-tail-distance asymptotic form. Eq.(\ref{psi})
represent Eq.(39) in Ref.\cite{KBS87}, deduced with the help of Eq.(39) in
Ref.\cite{KBS87} juxtaposed with Eq.(\ref{delta-r}). In turns, the
asymptotic {\em scatterer-collision} distribution function

\begin{equation}
D_{1}^{(s)}(n)=\Lambda _{s}[r(n)]\frac{dr}{dn}\varpropto \tau
_{c1}^{(w)}\left( n\tau _{c1}^{(w)}\right) ^{\frac{2}{z}-5}\text{, with }%
r(n)=n\tau _{c1}^{(w)}\gg 1\text{,}  \label{D1s-n}
\end{equation}
follows from Eq.(\ref{psi}) and the last relation in Eq.(\ref{n1-ran}).
Numerical analysis of Eq.(\ref{D1s-n}) is given in the reduced semi-log
coordinates in the left inset in Fig. 4. The best fitting for the proposed $%
D_{1}^{(s)}(n)$ with simulation data results in the derived dynamical
exponent $z_{1}=1.5$, which is the same obtained in Fig. 3 by the
variance-number analysis. This justifies the mechanism of long-range
L\'{e}vy jumps in the Bleher principal corridors. The universal
superdiffusion motion is also exposed by numerical analysis for the {\em %
wall-collision} distribution $\widetilde{D}_{1}^{(w)}(\widetilde{n})$, given
in the right plot in Fig. 4.

The velocity dispersive parameter $\sigma (R)$ was introduced in Ref.\cite
{KN00} to describe a disorder-to-order crossover in the weakly open SB in
terms of the normal-to-wall-velocity (${\rm v}_{\perp }$) pseudo-Gaussian
distribution function $g_{\sigma }(v)$, with $v={\rm v}_{\perp }/{\rm v}$.
The wall-collision statistics could be also introduced\cite{KN00} through
the velocity-dependent random number given by $n(v,t)=2vt/\tau _{c}^{(w)}$,
with the mean

\begin{equation}
n_{c\sigma }(t)=\int_{0}^{1}n(v,t)g_{\sigma }(v)dv=\frac{t}{\tau _{c}^{(w)}}%
\text{, and }g_{\sigma }(v)=\frac{1}{\sigma \sqrt{2\pi }}\frac{\exp [-(v-%
\frac{1}{2})^{2}/2\sigma ^{2}]}{%
\mathop{\rm erf}%
(1/2\sqrt{2}\sigma )}\text{.}  \label{nc-sigma}
\end{equation}
Here $%
\mathop{\rm erf}%
(x)$ is the standard error function. Two theoretical predictions were
proposed for the dispersive parameter, namely

\begin{equation}
\sigma (R)=\frac{\sqrt{1-2R/L}}{12\sqrt{5}(R_{2}/L)^{2}}\text{, for }R<R_{2}=%
\frac{L}{2}\text{,}  \label{sigma-VK}
\end{equation}
obtained in the simple ''quasi-chaotic'' ($\sigma \ll 1$) approximation\cite
{KN00} and\cite{Tese}

\begin{equation}
\sigma (R)=\sigma (0)\sqrt{\frac{4}{\pi }\frac{\arcsin (\frac{1-2R/L}{\sqrt{%
1+(1-2R/L)^{2}}})}{1-\pi (R/L)^{2}}}\thickapprox \sigma (0)\sqrt{\frac{1-2R/L%
}{1-\pi (R/L)^{2}}}\text{ ,}  \label{sigma-EV}
\end{equation}
which takes into consideration the long-living bouncing ball orbits and the
square-billiard data\cite{Tese} $\sigma (0)=0.29$. In the right inset in
Fig. 4, the simulation data for the weakly open SB is compared with the
theoretical predictions made for $\sigma (R)$. Remarkably, that similarly to 
$z_{1}(R)$, the late-time dispersion parameter is $R$-independent, within
the experimental error, and can be therefore characterized by $\sigma
_{1}(R)=0.14\pm 0.02$ for the universal diffusion regime.

\subsubsection{Finite Horizon Geometry}

In the case of LG with $R>L/2$, when all the Bleher corridors are closed
(see inset B in Fig.1), an application of the central limit theorem for
random-walk displacements ${\bf r}(t)$ proves\cite{BS81} its Gaussian
distribution. By taking into account the established in Eq.(\ref{delta-n})
relation between the random displacements and collision numbers, one may
expect the normal collision distribution, namely

\begin{equation}
D_{2}^{(w)}(n,t)=\frac{1}{\sqrt{2\pi \Delta ^{2}n_{c}(t)}}\exp \left[ -\frac{%
\left( n-n_{c}(t)\right) ^{2}}{2\Delta ^{2}n_{c}(t)}\right] \text{, for }%
R\geq L/2  \label{D-2}
\end{equation}
where $n_{c}$ and $\Delta ^{2}n_{c}$ are the standard mean and deviation of
the wall-collision number. Experimental justification of the asymptotic
Gaussian distributions $\widetilde{D}_{2}^{(w)}(\widetilde{n})$ $=$ $%
\widetilde{D}_{2}^{(s)}(\widetilde{n})=(1/2\pi )\exp (-\widetilde{n}^{2}/2)$%
, for both the wall and the scatterer collisions are given in {\bf Fig. 5}.
In general, our data on $\widetilde{D}_{2}^{(s)}(\widetilde{n})$ are
consistent with the first observation of the normal {\em scatterer-collision}
distribution reported in Ref.\cite{GG94}. Furthermore, our short-time
analysis provides evidence that the Gaussian distribution, associated with
chaotic dynamics, becomes steady at times $t_{obs}\gtrsim 50\tau _{c}$.

It is noteworthy that no true Gaussian distribution was achieved for the
late-time diffusion coefficient\cite{Ble92}, when $R$ tends to $L/2$ from
below. With the aim to clarify a decay process of spatial correlations
within this regime, we have analyzed the reduced fourth-order moments $\zeta
_{4}$ defined as 
\begin{equation}
\zeta _{4}(R)=\frac{\Delta ^{4}n_{c}(t)}{\Delta ^{2}n_{c}(t)}\text{, with }%
\frac{t}{\tau _{c}}\gg 1\text{,}  \label{ksi-m}
\end{equation}
and obtained with the help of Eq.(\ref{delta-n-k}). The true Gaussian
distribution prescripts $\zeta _{4}=3.$ Our temporal analysis given for the
case of $R/L=0.6$, results in $\zeta _{4}(t)=3\pm 0.01$ for the observation
times $t\gtrsim 70\tau _{c}$ . The distinct cases of the wall and scatter
collisions are described by the characteristic times $\tau _{c}$ given in
Eq.(\ref{tau-c}). An approximation towards the Gaussian distributions, with
the increasing of the disorder with $R/L$, is exposed in the insets in Fig.
5. One can see that a chaotic motion in the SB with $R>L/2$ is established
by both the normal collisions and by the normal diffusion. These universal
regimes are observed with a good precision, which is guaranteed by,
respectively, the higher-order central moment $\zeta _{4}=3.00\pm 0.04$ and
the diffusion exponent $z_{2}=2.0\pm 0.1$.

\section{CONCLUSION}

We have discussed the superdiffusive behavior of SB, as well as of the
dynamically equivalent periodic LG, in view of the interplay between the
linear and the circular boundary geometries. The late-time correlations,
driven by elastic reflections from the square walls and the dispersed disk,
are shown to be steady for a given geometry at observation times $30$-$%
50\tau _{c}(R)$. This justifies a consideration of the statistical
distributions for the random wall and scatterer collisions. Besides the
higher-order correlation effects, these collisions are characterized by
their mean collision numbers $n_{c}^{(w)}$ and $n_{c}^{(s)}$, respectively.
The relative collision numbers $n_{c}^{(w)}/n_{c}^{(s)}=\tau _{c}^{(s)}/\tau
_{c}^{(w)}=2L/\pi R$ follow from Eqs.(\ref{tau-c}) and (\ref{1/tau-c}).

When the scatterer is absent, the deterministic description results in the
ballistic motion given by the diffusion exponent $z_{0}=1$. The
wall-collision distribution is flat and asymmetric due to a sharp
contribution from the bouncing-ball orbits\cite{RC02} (see Fig. 2). At very
small scatterer radius, when the wall collisions predominate ($%
n_{c}^{(w)}\gg n_{c}^{(s)}$), the late-time distribution tends to a smooth
pseudo-Gaussian form, asymmetrically shifted from the center by the same
long-living bouncing-ball orbits (see Fig. 4). When the radius of scatterers
is relatively small, $R<R_{1}=L\sqrt{2}/4$, all the principal free-motion
corridors remain open in the corresponding LG, and the bouncing-ball
trajectories evolve freely within the diagonal and nondiagonal corridors.
This evolution, described by random walks driven by rare collisions with the
scatterers, is revealed through the long-distance L\'{e}vy jumps along the 
{\em diagonal} corridors. This stochastic motion is shown to be
radius-independent and characterized by the universal diffusion exponent $%
z_{1}=1.50\pm 0.05$, as well as by the constant pseudo-Gaussian
disorder-order parameter $\sigma _{1}=0.14\pm 0.02$. It seems plausible to
associate the underlying mechanism of the universal superdiffusion dynamics
with the critical behavior of spatial correlations\cite{MKW01}, with $%
z_{cr}=3/2$, described by the unbounded orbits moving in the diagonal Bleher
corridors. When the diagonal corridors become closed, one observes a
transient motion regime for $R_{1}\leq R<R_{2}$ and characterized by the
nonuniversal diffusion with $z_{1}<z(R)<2$. For the geometry with $R\geq
R_{2}=L/2$ , the scatterer overlaps the boundary walls and all the
free-motion corridors close. In this case, the disordered motion is due to
the bounded-orbit normal diffusion with $z_{2}=2$ and to the normal
collisions with $n_{c}^{(w)}\thickapprox n_{c}^{(s)}$. The bouncing-ball
motion disappear and the central symmetry of the collision distribution is
recovered. Despite of the fact that the normal correlations were earlier
expected for this billiard geometry from rigorous theory\cite{BS81} and
observed in numerical experiments\cite{GG94,RC02}, we have proven that both
the observed wall and scatterer collisions are true Gaussian.

Financial support of CNPq is acknowledged. 

\begin{center}
{\large \ }
\end{center}

\begin{figure}[tbp]
\end{figure}

Fig. 1. Sinai billiard and the corresponding periodic Lorentz Gas model for
different boundary geometries. {\em Case A}: reduction of a trajectory in
the LG lattice to the corresponding orbit in SB is shown through the
billiard wall-to-wall and wall-to-disk collision points $1,2,3,5$ and $4,6$,
respectively. {\em Case B}: infinite horizon geometry for $0<R<L\sqrt{2}/4$.
Examples of the principal free-motion corridors by Bleher\cite{Ble92}: the
nondiagonal (a) and diagonal (b) corridors. {\em Case C}: finite horizon
geometry, $R\geq L/2$. {\em Case D}: reduction of the unbounded trajectory
of launching angle $\theta $ in LG without scatterers to the corresponding
orbit in the square billiard ($R=0$).

.

\begin{figure}[tbp]
\end{figure}

Fig. 2. Analyses of the wall-collision dynamics in square billiard. Points
correspond to simulation data on the collision function observed at $%
t_{obs}=100\tau _{c0}$ and on the temporal evolution for the
root-mean-square collision-number deviation, with $\tau _{c0}=\pi /4$. Solid
lines are the corresponding theoretical predictions given by Eq.(\ref{D-0})
and Eq.(\ref{var}).

.

\begin{figure}[tbp]
\end{figure}

Fig. 3. Diffusion dynamic exponent against the reduced radius of the
scatterer disk in Sinai billiard. Points are simulation data derived from
the observed billiard-wall collisions through the standard deviation (\ref
{delta-n-z}) and the estimated characteristic time $\tau _{c}^{(w)}(R)$
given in Eq.(\ref{tau-c}). Inserts: temporal evolution of the standard
wall-collision-number deviation for different infinite ($R<L/2$, upper
inset) and finite ( $R>L/2$, lower inset) horizon geometries, in the log-log
coordinates. The solid lines are the best linear fitting of the simulation
data.

.

\begin{figure}[tbp]
\end{figure}

Fig. 4. Collision distributions against collision numbers in Sinai Billiard
with the infinite horizon geometry ($R<L/2$). Points are simulation data for
the scatterer (left plot) and the wall (right plot) collisions represented
in the reduced coordinates defined in Eq.(\ref{D-red}). The solid lines are
the Gaussian distributions $\widetilde{D}_{2}(\widetilde{n})=(1/2\pi )\exp (-%
\widetilde{n}^{2}/2)$. Left inset: analysis of the late-time
scatter-collision function for long-jump-scatterer collisions. Points are
simulation data for the case $R=0.25L$ and $t_{obs}=200\tau _{c}^{(w)}$.
Solid line is the best fitting with the prediction given in Eq.(\ref{D1s-n}%
), with $\widetilde{D}_{1}^{(s)}(\widetilde{n})=0.5|\widetilde{n}|^{-11/3}$.
Right inset: velocity-dispersive parameter against reduced scatterer radius.
Points are simulation asymptotic data\cite{Tese} for the weakly open SB;
dashed and solid lines are theoretical predictions given in Eqs. (\ref
{sigma-VK}) and (\ref{sigma-EV}), respectively.

.

\begin{figure}[tbp]
\end{figure}

Fig. 5. Observation of the Gaussian collision statistics in Sinai Billiard
with $R>L/2$. Points are simulation data for the scatterer (left plot) and
the wall (right plot) collisions observed at $t_{obs}=200\tau _{c}$ for
distinct geometries indicated in the legend. The lines are the same as in
Fig. 4. In the left and right inserts, the analyses of the fourth moment
(defined in Eq.(\ref{ksi-m})) for the scatterer and wall collisions are
given at $t_{obs}=200\tau _{c}$, respectively. Points are simulation data
and the dashed line corresponds to $\zeta _{4}=3$.

\end{document}